\documentclass[aps,prl,twocolumn,groupedaddress]{revtex4}

\usepackage{graphicx}
\usepackage{amsmath}
\graphicspath{{Figures/}}

\newcommand{\ket}[1]{\left|#1\right\rangle}

\begin{document}


\title{Direct Observation of Coherent Population Trapping in a Superconducting Artificial Atom}


\author{William R. Kelly}
\email{wkelly@bbn.com}
\author{Zachary Dutton}
\email{zdutton@bbn.com}
\author{John Schlafer}
\author{Bhaskar Mookerji}
\author{Thomas A. Ohki}
\email{tohki@bbn.com}
\affiliation{Raytheon BBN Technologies, Cambridge, MA 02138, USA}
\author{Jeffrey S. Kline}
\author{David P. Pappas}
\affiliation{National Institute of Standards and Technology, Boulder, CO 80305, USA}


\date{\today}

\begin{abstract}
The phenomenon of Coherent Population Trapping (CPT) of an atom (or solid state ``artificial atom''), and the associated effect of Electromagnetically Induced Transparency (EIT), are clear demonstrations of quantum interference due to coherence in multi-level quantum systems.  We report observation of CPT in a superconducting phase qubit by simultaneously driving two coherent transitions in a $\Lambda$-type configuration, utilizing the three lowest lying levels of a local minimum of a phase qubit.  We observe $60(\pm 7)\%$ suppression of excited state population under conditions of CPT resonance.   We present data and matching theoretical simulations showing the development of CPT in time.  Finally, we used the observed time dependence of the excited state population to characterize quantum dephasing times of the system.
\end{abstract}


\maketitle


Coherent population trapping (CPT) is a physical effect in which a quantum system is driven into a coherent superposition of two long-lived states by a combination of two optical fields.  These fields couple each of the long-lived states to a common, shorter-lived excited state.  CPT has been observed for some time in atomic systems \cite{GrayCPT}.  CPT refers to the dynamics of the atom while the associated effect on the optical fields is referred to as electromagnetically induced transparency (EIT) \cite{harris}.  EIT has been shown to give rise to a number of interesting effects, including slow light \cite{SlowLightHau,SlowLightScully,SlowLightBudker}, low light level non-linear optics \cite{NLOHarris,NLOImamoglu}, and single photon quantum storage and memory \cite{SinglePhotonStorageLukin, SinglePhotonStorageKimble, SinglePhotonStorageKuzmich}.  There have been impressive demonstrations in quantum wells \cite{QuantumWellEITPhillips,QuantumWellEITWang} and more recently in quantum dots \cite{QuantumDotEIT}.

In \cite{DuttonEIT} it was proposed to apply these concepts to the microwave, rather than optical, domain in superconducting quantum circuits.  This step is interesting for two reasons.  First, it applies CPT and EIT and related nonlinear optical effects to superconducting qubits, which have shown great promise for future quantum information processing systems \cite{DiCarlo2Qubit,MartinisBellIneq}.  Second, it also involves observing these effects in a completely different region of the electromagnetic spectrum.  Some experimental steps in this direction were taken in \cite{BaurATSplitting,SimmondsATSplitting}, in which the related effect of Autler-Townes (AT) splitting \cite{AutlerTownes} was observed.

In this letter, we present direct experimental observation of Coherent Population Trapping (CPT) in a three level $\Lambda$ configuration superconducting system in which two long-lived levels ($\ket 0$ and $\ket 1$) are coupled to a short-lived higher-lying level ($\ket 2$).  By resonantly driving two transitions of this system with a two-tone microwave field, we observe the creation of a coherent superposition state of the lower lying levels, which is known as a ``dark'' state and is central to the observation of CPT \cite{harris}.   We observe the probe resonance in the absence of CPT conditions, and a $60(\pm 7)\%$ suppression of this resonant excitation under conditions of CPT.   Furthermore, we observe the time dynamics of the development of the CPT resonance and use the observed time dependence to extract decoherence rates, as proposed in \cite{DuttonEIT}.  Our experimental results are corroborated by theoretical simulations of the density matrix evolution of the system.

\begin{figure}
    \includegraphics[width=8.5 cm]{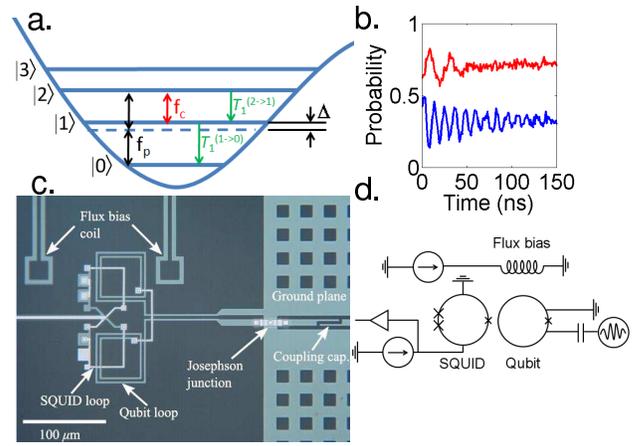}%
    \caption{(a) Single well anharmonic potential shown with four energy levels. Arrows illustrate the dominant driven transitions and decoherence channels in our system three level $\Lambda$ system.  (b) Rabi oscillation of the $f_{01}$ (blue) and $f_{02}$ (red) transitions are shown for potential with $f_{01} = 6.7$~GHz.  The two curves are offset for clarity.  (c) A micrograph of the measured qubit.  The Josephson junctions in the circuit are epitaxially grown Rhenium with crystalline Al$_2$O$_3$ tunneling barriers. (d) Circuit diagram of qubit, SQUID, and flux bias line.  The qubit is controlled via the flux bias line and the capacitively coupled microwave line.  The qubit state is probed by current biasing the SQUID and measuring the resulting voltage.\label{ExpSystem}}
\end{figure}

In our implementation, the three states of our $\Lambda$ system are the three lowest levels of a metastable phase qubit potential well \cite{Mart} (Fig.~\ref{ExpSystem}(a)).  This device exhibits an anharmonic potential in which the transition frequencies, $f_{01}$ and $f_{12}$, differ by a few percent.  The design of the qubit seen in Fig. \ref{ExpSystem}(c,d) is identical to the devices explored in \cite{Kline}.  The fabrication of the junctions was targeted to yield a flux tunable transition frequency in the 5-10 GHz range.  For the CPT experiment we biased the qubit to have $f_{01} = 6.205$~GHz which makes $f_{12} = 5.865$~GHz.  This corresponds to a shallow qubit potential that supports approximately 8 metastable bound states.  For the sake of visual simplicity these other levels are not include in Fig. \ref{ExpSystem}(a).  We choose this flux bias to make the anharmonicity parameter $\Delta \equiv (f_{01}-f_{12})/2 = 170$~MHz sufficiently large and to create a strong damping of $\ket 2$ which is advantageous for observing CPT.

When a microwave field is applied at the frequency $f_{01}$ we obtain Rabi oscillations with a period determined by the Rabi frequency $\Omega^{(01)}$, proportional to the field envelope amplitude.  To create a $\Lambda$ system, we couple $| 0 \rangle$ to $| 2 \rangle$ via a two-photon virtual transition with a {\it probe} microwave field near frequency $f_{02}/2 = 6.035$~GHz.  In this virtual excitation scheme the single photon energy is significantly detuned from $f_{01}$.  The detuning is determined by the anharmonicity parameter $\Delta$.  The resonant Rabi frequency associated with the $|1 \rangle \rightarrow | 2 \rangle$ transition is $\Omega_p^{(12)}=\sqrt{2} \Omega_p^{(01)}$ and for sufficiently small microwave field amplitude ($\Omega_p^{(01)},\Omega_p^{(12)} \ll \Delta$),  the effective two-photon coupling amplitude is $ A_p =  \left(\Omega_p^{(01)}\right)^*\Omega_p^{(12)}/2\Delta $ \cite{Dutta}.  The accuracy of these expressions was confirmed in \cite{Dutta} where it was noted that this two-photon virtual transition amplitude quantitatively explained the observed coupling rates.  It was also noted there that the two-photon virtual process dominates the direct two-photon $| 0 \rangle \leftrightarrow | 2 \rangle$ process.   Contrary to the spectroscopic results in \cite{Dutta}, we observe only the $f_{01}$ and $f_{02}$ transitions and none of the higher level transitions observed in \cite{Dutta}.  In our experiment we choose $\Omega_p^{(01)} \approx (2 \pi) 48$~MHz, giving $A_p=(2 \pi) 6.8$~MHz.

Rabi oscillations shown in Figure 1(b) were measured with a qubit bias that exhibited longer coherence times [23] than the bias used for the CPT experiment. This demonstrates coherence of both the $\ket 0 \leftrightarrow \ket 1$ and $\ket 0 \leftrightarrow \ket 2$ transitions.  The $\ket 0 \leftrightarrow \ket 2$ oscillations are slower than the $\ket 0 \leftrightarrow \ket 1$ oscillations due to the smaller effective Rabi frequency, and more damped due to the faster relaxation of state $| 2 \rangle$ \cite{MartinisFockStateDecay}.  We note that in this scheme there is some unwanted population excitation of the intermediate level $|\Omega_p^{(01)}|^2/2 \Delta^2 \approx 3\%$.  This process is accounted for in our theoretical analysis and it effects our data quantitatively but does not prevent CPT from clearly being observed.

To suppress the probe transition process via CPT, we apply a second {\it coupling} microwave field at frequency $f_{12}$. The amplitude of this transition is given by $A_c = \Omega_c^{(12)}$, where $\Omega_c^{(12)}$ is the resonant Rabi frequency.  When both fields are applied simultaneously, the system is ideally driven from the ground state $\ket 0$ into the ``dark'' state superposition $|\psi \rangle = (A_c| 0 \rangle - A_p |1 \rangle)/A$, where $A=\sqrt{|A_p|^2+|A_c|^2}$.  In this superposition, the excitation amplitudes for each of the individual probe and coupling transitions are equal and opposite, resulting in a quantum interference which suppresses the excitation of state $|2 \rangle$.  

In order to initialize the system into a well defined state, we let the qubit relax to the state $\ket 0$. To observe CPT we then apply a two-tone microwave field at frequencies $f_p$ and $f_c$, with independently controlled power at each frequency, and duration $t_0$ using a pulsing scheme similar to~\cite{MartinisSingleQubitGate}.  Our microwave pulses were generated by sending trapezoidal envelope pulses through a 1~GHz low pass filter and into the IF port of a microwave mixer.  Analysis of Rabi and $T_1$ data showed that we had independent control over the amplitude and duration of these pulses.  After the time $t_0$, the qubit state is destructively measured with a calibrated pulse that probes the population of the state $\ket{2}$ only ($P_2$) \cite{MeasEndnote}.  We repeat this procedure many times at each $f_p$ and $f_c$ to get statistics, and swept these frequencies about the resonances $f_{02}/2$ and $f_{12}$.

\begin{figure}
    \includegraphics[width=8.5 cm]{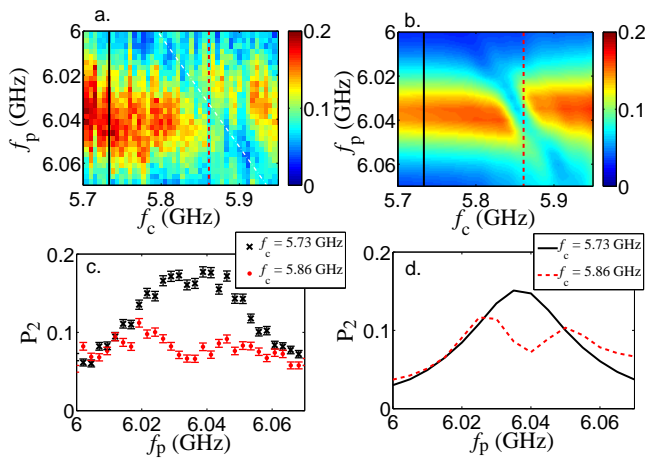}%
    \caption{Demonstration of CPT in a phase qubit.  (a) Population of state $\ket{2}$ ($P_2$) measured after a two tone microwave pulse with a duration of $t_0=30$~ns and frequencies $f_p$ and $f_c$.  Each population measurement was performed by counting the number of SQUID switching events out of 4,000 trials and accounting for 80\% measurement fidelity and 3\% background counts.  The power levels for the probe and coupling fields were -6.5~dBm and -15~dBm at the source, respectively.  The white dot-dashed line shows the two CPT resonance $2f_p-f_c=f_{01}$  (b) Theoretical data obtained by numerically solving the four level master equation (see text).   We obtained the best match with experimental data by setting the resonant Rabi frequencies for these powers to be $\Omega_p^{(01)}=(2 \pi)48$~MHz, $\Omega_c^{(12)}=(2 \pi)32$~MHz.   These values yield $A_p=|\Omega_p^{(01)}|^2/2 \Delta = 6.8$~MHz, $A_c=\Omega_c^{(12)}=32$~MHz. (c-d) Cross sections of experimental data (with statistical error bars) (c) and theoretical data (d) taken at the values of $f_c$ indicated by the solid and dashed lines in (a) and (b).  The cuts show the one-photon resonance when $f_c$ is off-resonance and show the suppression of $P_2$  when $f_c=5.86$ GHz.\label{FreqVsFreq}}
\end{figure}

Figure~\ref{FreqVsFreq}(a) presents our results showing the CPT resonance and Fig.~\ref{FreqVsFreq}(c) shows cuts at two values of $f_c$.   At $f_c=5.73$~GHz, far from the coupling resonance $f_{12}$, we observe a probe resonance at $f_p=6.035$~GHz with width of approximately 40~MHz.  This is primarily broadened by the linewidth of the $\ket 0 \leftrightarrow \ket 2$ transition, $\gamma_{02} = 1/2T_1^{(2\rightarrow 1)}+1/T_{\phi}^{(02)}$.  No CPT is observed here as the dark state is quickly dephased.  A similar feature is obtained in the absence of a coupling field.  By contrast, a cut at $f_c=5.86$~GHz reveals a $60(\pm 7)\%$ suppression of $P_2$ relative to the off-resonant case.  The CPT contrast and uncertainty was extracted from a parabolic fit of both curves in Fig.~\ref{FreqVsFreq}(c).  A salient feature observed in Fig.~\ref{FreqVsFreq}(a) is the diagonal trench of suppressed $P_2$ at positions corresponding to the CPT resonance $2 f_p - f_c = f_{01}$ (white dot-dashed line in Fig.~\ref{FreqVsFreq}(a)).

We compare our experimental results with theoretical simulations in which we propagate the density matrix equations in time with a Runge-Kutta algorithm.  In the model we include the first four levels of the phase qubit potential well and dipole coupling on $| 0 \rangle \leftrightarrow |1 \rangle$,  $| 1 \rangle \leftrightarrow |2 \rangle$, and $| 2 \rangle \leftrightarrow |3 \rangle$ from both applied fields.  The relative dipole moments for these transitions are $1:\sqrt{2}:\sqrt{3}$.  The presence of the state $\ket 3$ was seen to have a negligible effect on our simulation results, as was expected.  The population of state $\ket{3}$ was found to be  $< 0.5\%$ for all of our simulations.  The overall calibration between Rabi frequency couplings $\Omega_{p,c}^{(01)}$ and microwave powers was fitted as a free parameter.  Since the relative phase of the probe and coupling field varied randomly from shot to shot in our experiment, our theoretical results were averaged over all possible phase differences of the microwave fields.  The dipole moment for direct $| 0 \rangle \leftrightarrow | 2 \rangle$ transitions is much smaller \cite{Dutta} and furthermore is far detuned from both microwave fields. Population decay was included according to independent measurements of the energy relaxation times, $T_1^{(1 \rightarrow 0)}=108$~ns and $T_1^{(2 \rightarrow 1)}=77$~ns.  We extracted the dephasing time $T_\phi^{(02)}=6(\pm 1)$~ns from the width of several off resonant cuts near $f_c=5.73$~GHz.

Results of our theoretical simulations are presented in Fig.~\ref{FreqVsFreq}(b).  We see good agreement with the experimental data in Fig.~\ref{FreqVsFreq}(a), with the theory clearly showing the broad horizontal stripe corresponding to the probe resonance and the diagonal trough due to CPT. A more quantitative comparison is seen by examining the cuts in Fig.~\ref{FreqVsFreq}(c) and (d).  By fitting data near the CPT resonance at $f_c = 5.865$~GHz, we found best agreement for dephasing times $T_\phi^{(01)}=30$~ns and $T_{\phi}^{(12)}=30$~ns.  Rabi data taken for this potential was consistent with this value of $T_\phi^{(01)}$.  In the limit of a perfect three-level $\Lambda$ system and an infinitely long lived coherence $\gamma_{01}=1/2T_1^{(1\rightarrow 0)}+1/T_{\phi}^{(01)}=0$, the CPT suppression under conditions of EIT resonance ($2 f_p-f_c=f_{01}$) would be $100\%$.  The resonant $P_2$ suppression predicted in the theoretical simulation is $52\%$, in fairly good agreement with the observation of $60(\pm 7)\%$.  Analysis of the theoretical results reveals that the finite probability of excitation is primarily due to dephasing of the $\ket{0} \leftrightarrow \ket{1}$ transition, but also partly due to undesired off-resonant coupling \cite{DuttonEIT2}.  The most dominant such coupling is due to the probe field coupling the $| 1 \rangle \leftrightarrow |2 \rangle$ transition.

We note that our configuration is distinct from that employed in the recent observation of Autler-Townes splitting in a superconducting phase qubit \cite{SimmondsATSplitting}, where a ladder scheme was used and population was suppressed in the intermediate level $| 1 \rangle$.  This leads to two important qualitative differences.  First, in contrast to \cite{SimmondsATSplitting}, we directly measure the expected population suppression (which in our case is of $|2 \rangle$).  Second, the fact that we create a dark state of the two longest lived states in the system allows us to unambiguously observe CPT, rather than the closely related phenomena of Autler-Townes (AT) splitting \cite{SimmondsATSplitting,BaurATSplitting}.  In AT, the coupling field dresses the system, splitting the probe resonance, of width $\gamma_{02}$, into two distinct resonances, spaced by the coupling field amplitude $A_c \gg \gamma_{02}$.  In this regime, the suppression of $P_2$ at the probe resonance can be attributed to large detuning ($\approx A_c/2$) from the dressed resonance peaks.  By contrast, when $A_c < \gamma_{02}$ the resonances still largely overlap and the suppression of $P_2$ can only be explained by a quantum inteference of the excitation processes.  In our experiment, we estimate $A_c=(2 \pi)32$~MHz, and $\gamma_{02} = 27.5$~MHz, which would correspond to only $1 - 2 \gamma_{02}(\gamma_{02}+\gamma_{01})/\left((\gamma_{02}+\gamma_{01})^2+\Omega_c^2\right) = 17(\pm 13)\% $ suppression of $P_2$ attributable purely to the AT splitting.  The uncertainty in this quantity comes from the uncertainty in the value of $T_\phi^{(02)}$.  Using this same model the contrast including quantum interference is predicted to be $62\%$, which is consistent with our experimental results.

\begin{figure}
    \includegraphics[width=8.0 cm]{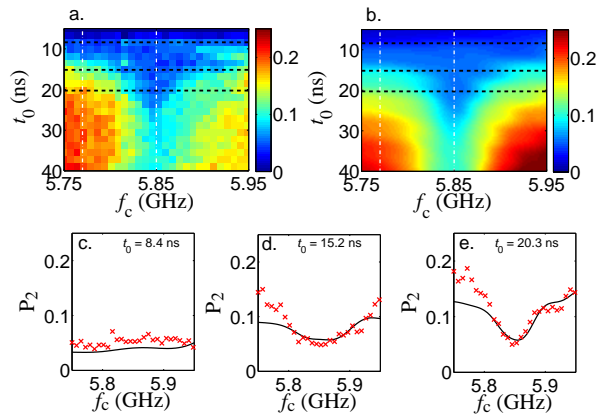}%
    \caption{Evolution of EIT as a function of microwave pulse duration.  In this data set $f_{01}=6.19$~GHz, $f_{12}=5.85$~GHz, and $f_{02}/2=6.02$~GHz, 15~MHz shifted from the data set in Fig.~\ref{FreqVsFreq}.  $f_p$ was held fixed at 6.0158~GHz, essentially on resonance with the $\ket 0\leftrightarrow \ket 2$ two-photon transition.  Each population measurement was an average of 5,000 trials.  (a) Experimental measurement of $P_2$, and (b) theoretical calculation of $P_2$.  The theoretical resonant Rabi frequencies were $\Omega_p^{(01)}=(2 \pi) 48$~MHz and $\Omega_c^{(12)}=(2 \pi) 35$~MHz.  (c)-(e) Cross sections of experimental data (symbols) and theory data (line) indicated by black dashed lines on (a) and (b).  Statistical error bars are approximately equal to the symbol size.  White dot-dashed lines indicate cross sections shown in Fig.~\ref{TimeDependence}. \label{TimeVsFreq}}
\end{figure}

To further understand CPT and EIT in our system, we varied the duration $t_0$ of the two microwave fields.  Figure~\ref{TimeVsFreq}(a) presents measurements of $P_2$ as a function of $t_0$ and $f_c$ keeping the probe field on the two-photon resonance $f_p=f_{02}/2$.  The results show the excitation into $|2 \rangle$ is slowed when the coupling field is at the CPT resonance $f_c=f_{12}$.  Figure~\ref{TimeVsFreq}(c-e) shows cuts in $f_c$ at several different times.  Note that the probe is always on, resonantly exciting the $| 0 \rangle \rightarrow | 2 \rangle$ transition.  As time progresses the CPT feature narrows and deepens until the maximum suppression ($\approx 55 \%$) is reached at about $20$ ns.  After this time, contrast is degraded by the decoherence rate $\gamma_{01}$ and in the limit of longer times, the EIT notch disappears.

\begin{figure}
    \includegraphics[width=8.0 cm]{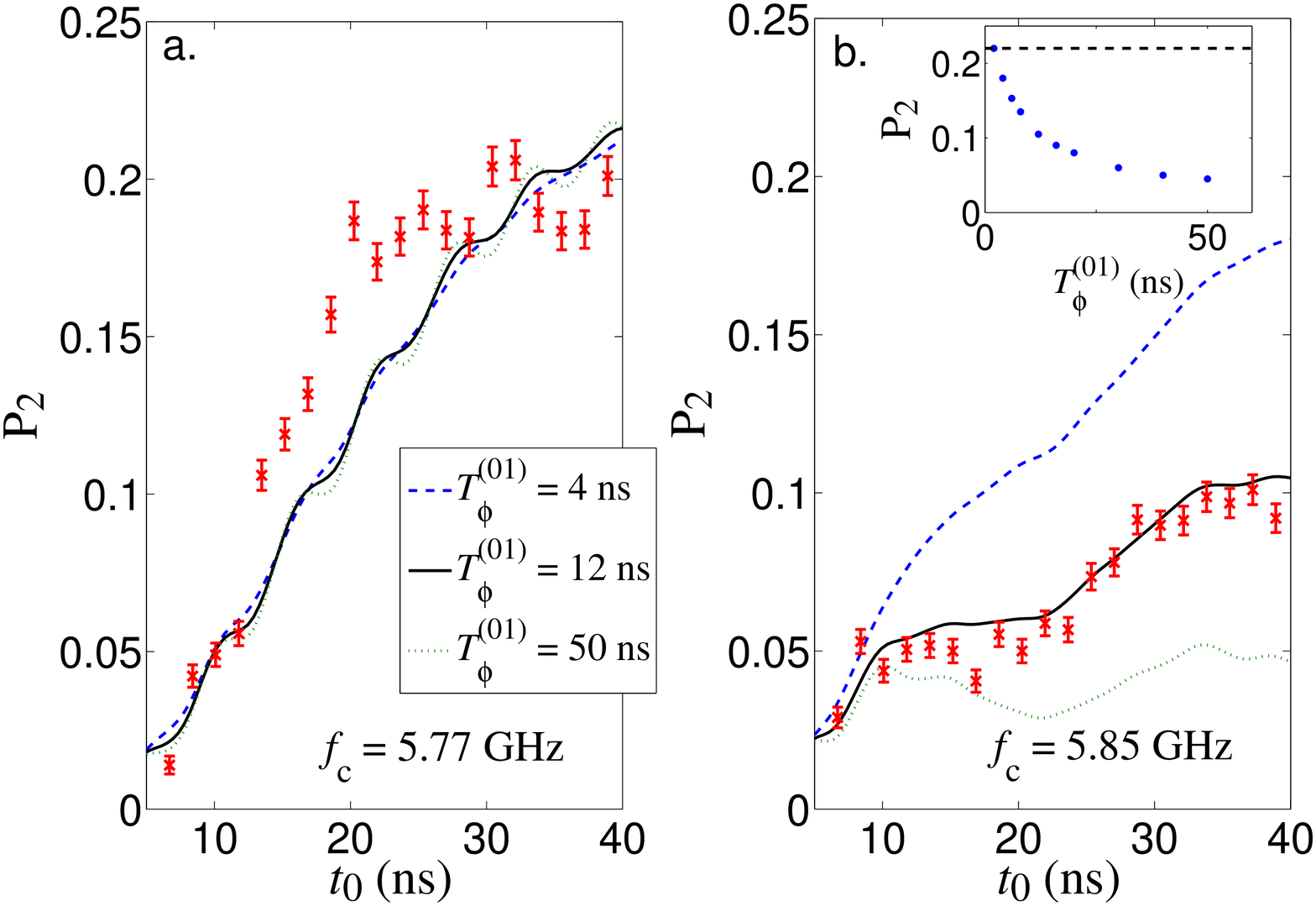}%
    \caption{Cuts in time of the experimental data (with statistical error bars) presented in Fig.~\ref{TimeVsFreq}(indicated by white dot-dashed lines in a,b) at two different frequencies $f_c$ corresponding to (a) off CPT resonance $f_{12}-f_c=80$ MHz and (b) on CPT resonance $f_{12}-f_c=0$ MHz.  Also shown are theoretical simulations of the evolution of $P_2$ over time shown for three different values of $T_\phi^{(01)}$ with other parameters the same as in Fig.~\ref{TimeVsFreq}.  (inset) $P_2$ for $f_c=f_{12}$ for $t_0=40$ ns plotted as a function of $T_\phi^{(01)}$.  The black dashed line shows the population off resonance ($f_{12}-f_{c}=80$ MHz).\label{TimeDependence}}
\end{figure}

Since the excitation of $P_2$ is driven by $\gamma_{01}$, it was previously proposed \cite{DuttonEIT}  to extract this decoherence rate by measuring loss in a CPT experiment.  We realized this idea by taking a cut of the data in Fig.~\ref{TimeVsFreq}(a) both off ($f_c=5.77$~GHz) and on ($f_p=5.85$~GHz) CPT resonance (Fig.~\ref{TimeDependence}(a,b)).  The theoretical data from Fig.~\ref{TimeVsFreq} are then presented as solid black lines.  We adjusted the various dephasing times and energy level frequencies in the theory to best fit the experimental data and found the best agreement for $T_\phi^{(01)}=12$~ns, $T_{\phi}^{(02)}=20$~ns , and $T_{\phi}^{(12)}=20$~ns \cite{TLSEndnote}.

To understand how sensitive a probe this measurement is of $T_\phi^{(01)}$ we performed several simulations in which we varied this parameter.  The inset in Fig.~\ref{TimeDependence}(b) shows $P_2$ at $t_0=40$~ns as a function of $T_\phi^{(01)}$.  We see that in the regime $(4~\mathrm{ns} < T_\phi^{(01)} < 30~\mathrm{ns})$ we are able to measure $T_\phi^{(01)}$.  At longer $T_\phi^{(01)}$ the population asymptotically approached a level $P_2 \approx 0.05$ which is dominated by off-resonant coupling effects \cite{DuttonEIT2}.  Performing our experiment in a regime with  greater anharmonicity would suppress this effect, allowing measurement of longer $T_\phi^{(01)}$ and, in the event of obtaining larger coherence times, greater CPT suppression of $P_2$.

In summary, we have directly observed CPT in a three-level superconducting phase qubit and seen suppression of the excited level population that is clearly attributable to quantum interference.  We did this by creating a $\Lambda$ configuration using a two-photon virtual probe transition and monitoring the population of $\ket 2$.  We have matched our results with theoretical simulations of the full density matrix equations, including expected decoherence and cross-talk effects.  We have furthermore observed the development of the characteristic EIT notch in time and used these time dynamics to characterize the decoherence time of the $\ket{0} \leftrightarrow \ket{1}$ transition.

These results are of fundamental interest as a clear observation of a quantum interference at a macroscopic scale.  Furthermore, they are directly pertinent to future investigations of low light level nonlinear optics in the microwave domain and quantum information applications utilizing superconducting quantum circuits.

\begin{acknowledgments}
We acknowledge generous internal support from BBN Technologies to carry out this work. We thank Jon Habif for the suggestion to try this experiment.  The fabrication of the device was funded by the US government and IARPA.
\end{acknowledgments}

\bibliography{EIT_bibliography}

\begin{thebibliography}{27}
\bibitem[{Mea()}]{MeasEndnote}
\bibinfo{note}{Using theoretical simulations we extracted a measurement
  contrast of $80 \%$, and background of $3 \%$ in our data. All presented data
  have been processed to give $P_2$ in accordance with this analysis.}
\bibitem[{TLS()}]{TLSEndnote}
\bibinfo{note}{Note that the dephasing times are different in
  Fig.~\ref{FreqVsFreq} and Fig.~\ref{TimeVsFreq}. We speculate that this
  difference could be attributable to shifts in the frequencies of two-level
  fluctuators between runs.}
\end{thebibliography}

\end{document}